\documentclass[twocolumn,prb,aps,amsmath,amssymb,superscriptaddress]{revtex4}
\usepackage{graphicx}
\usepackage{bm}

\renewcommand\Re{\operatorname{Re}}

\begin{document}
\title{K-edge X-ray absorption spectra in transition metal oxides
  beyond the single particle \\approximation: shake-up many body
  effects.}

\author{M. Calandra}
\affiliation{Institut de Min\'eralogie et de Physique des Milieux condens\'es, Universit\'{e} Paris 6, CNRS, 4 Place Jussieu, 7504 Paris, France}
\author{J. P. Rueff}
\affiliation{Synchrotron SOLEIL, L'Orme des Merisiers, Saint-Aubin, BP~48, 91192 Gif-sur-Yvette Cedex, France}  
\affiliation{Laboratoire de Chimie Physique--Mati\`ere et Rayonnement, CNRS-UMR~7614, Universit\'{e} Pierre et Marie Curie, F-75005 Paris, France} 
\author{C. Gougoussis}
\affiliation{Institut de Min\'eralogie et de Physique des Milieux condens\'es, Universit\'{e} Paris 6, CNRS, 4 Place Jussieu, 7504 Paris, France}
\author{D. C\'{e}olin}
\affiliation{Synchrotron SOLEIL, L'Orme des Merisiers, Saint-Aubin, BP~48, 91192 Gif-sur-Yvette Cedex, France}  
\author{M. Gorgoi}
\affiliation{Berliner Elektronenspeicherring-Gesellschaft f\"{u}r
  Synchrotronstrahlung m.b.H., Albert-Einstein-Strasse 15, D-12489
  Berlin, Germany}
\author{S.~Benedetti}
\affiliation{Centro S3, Istituto Nanoscienze-CNR, via Campi 213/a 41125, Modena, Italy} 
\author{P. Torelli}
\affiliation{CNR-Istituto Officina dei Materiali, Laboratorio TASC in Area Science Park, S.S. 14 km 163.5, Basovizza, 34149 Trieste, Italy}
\author{A. Shukla}
\affiliation{Institut de Min\'eralogie et de Physique des Milieux condens\'es, Universit\'{e} Paris 6, CNRS, 4 Place Jussieu, 7504 Paris, France}
\author{D. Chandesris}
\affiliation{Laboratoire de Physique des Solides, CNRS-UMR 8502, Universit\'{e} Paris-Sud, F-91405 Orsay, France}  
\author{Ch. Brouder}
\affiliation{Institut de Min\'eralogie et de Physique des Milieux condens\'es, Universit\'{e} Paris 6, CNRS, 4 Place Jussieu, 7504 Paris, France}
\date{\today}

\begin{abstract}
The near edge structure (XANES) in K-edge X-ray absorption spectroscopy (XAS) is a widely used tool for studying electronic and local structure in materials. The precise interpretation of these spectra with the help of calculations is hence of prime importance, especially for the study of correlated materials which have a complicated electronic structure per se. The single particle approach, for example, has generally limited itself to the dominant dipolar cross-section. It has long been known however that effects beyond this approach should be taken into account, both due to the inadequacy of such calculations when compared to experiment and the presence of shake-up many-body satellites in core-level photoemission spectra of correlated materials. This effect should manifest itself in XANES spectra and the question is firstly how to account for it theoretically and secondly how to verify it experimentally. 
By using state-of-the-art first principles electronic structure calculations and 1s photoemission measurements we demonstrate that shake-up many-body effects are present in K-edge XAS dipolar spectra of NiO, CoO and CuO at all energy scales. We show that shake-up effects can be included in K-edge XAS spectra in a simple way by convoluting the single-particle first-principles calculations including core-hole effects with the 1s photoemission spectra. We thus describe all features appearing in the XAS dipolar cross-section of NiO and CoO and obtain a dramatic improvement with respect to the single-particle calculation in CuO.
These materials being prototype correlated magnetic oxides, our work points to the presence of shake-up effects in K-edge XANES of most correlated transition metal compounds and shows how to account for them, paving the way to a precise understanding of their electronic structure.
\end{abstract}
\maketitle

\section{Introduction}

Low-energy excitations of correlated materials are strongly influenced by
electron-electron interaction effects beyond the single particle
approximation. This is particularly evident in
core-level photoemission spectra (XPS) of transition metal compounds
where the occurrence of many-body satellites is well documented
(for a review see Ref. \onlinecite{Groot_Kotani_Book}).
As for x-ray absorption, interpretation of the dipolar K-edge XAS
cross-section heavily relies on standard single-particle first principles
calculations ~\cite{Gougoussis09a,Taillefumier,Hebert_Wien2k,Joly} that neglect shake-up excitations.
Since dipolar L$_{2,3}$ XAS mostly samples d-states of the absorbing atom which are more prone to effects of correlation
than p-states, one normally assumes that shake-up effects are
visible mostly in L$_{2,3}$ XAS and not in K XAS. However a recent work~\cite{Gougoussis09b} shows that in NiO 
the single particle dipolar K-edge spectrum misses some near-edge and
far-edge features present in the experimental measured one.

We concentrate on shake-up many body excitations arising
from a valence electron excitation following the creation of a core
hole by the incident x-ray~\cite{Agren92}. 
In the past, the description of shake-up effects in core-hole photoemission spectra
has been investigated in the framework of quantum-chemical
calculations ~\cite{Carravetta-12}, by using approaches
based on model hamiltonians \cite{Veenendal93,Groot_Kotani_Book,
Taguchi} or first-principles modified approaches ~\cite{Takahashi-12}.
The occurrence of these effects in XPS is well established but they have also been shown to occur in M$_{4,5}$ edges of
mixed-valent compounds \cite{Gunnarsson} and in L$_{2,3}$ X-ray absorption
spectra (XAS) of transition metals and rare earths compounds
\cite{Hammoud, Malterre} and were proposed as a
possible explanation of the double peak structure in dipolar
K-edge XAS of high T$_c$ cuprates \cite{tolentino_PhysRevB.45.8091} and copper compounds in
general \cite{Bair_PhysRevB.22.2767}. However this attribution was
questioned in Ref. \cite{Kosugi_PhysRevB.41.131, Gougoussis09a} 
and the double peak structure was suggested to be single
particle in origin.

Nailing down the importance of these effects has been difficult due to complications related to many-body calculations but also to the paucity of experimental 1s photoemission spectra. In this work, following earlier suggestions, we demonstrate that shake-up manybody effects can be included in a simple way
in K-edge XAS spectra by convoluting the single particle first principles calculations 
with experimental 1s photoemission spectra, some of which we have freshly measured.
We show that this procedure explains all features in K-edge XAS spectra of NiO and CoO and
strongly improves the agreement with experimental data in CuO.
Our work points out the relevance of these effects in K-edge dipolar
XAS of all compounds displaying multiple structures in photoemission
spectra.

The structure of the paper is the following. In section 
\ref{sec:theory}, following
Ref \onlinecite{OhtakaRMP}, we briefly sketch the 
demonstration of the convolution formula relating
the many body XAS cross section to the single particle
one via the photoemission spectrum.  We  then present 
experimental
details concerning our measured photoemission spectra in 
sec. \ref{sec:experiment}. The experimental results
and the theoretical undestanding are presented in sec. \ref{sec:results}.

\section{Theory}

\subsection{Shake-up theory}\label{sec:theory}
Shake up satellites are many-body peaks present in 
core-electron spectra. They originate from a valence
electron excitation following the creation of a core
hole by the incident x-ray~\cite{Agren92}. 
Quantum chemical calculations of shake-up
satellites have been recently reviewed by
Carravetta and {\AA}gren~\cite{Carravetta-12}.

An electric dipole transition between two
Slater determinants built from the same set
of orbitals do not allow for shake-up satellites.
Indeed, the orthogonality of orbitals allows
for only one transition from the core level
to the empty one. Therefore, a shake-up can
only be obtained by describing the (initial)
state with a linear combination of Slater
determinants or by using different orbitals
for the initial and final 
determinants~\cite{Martin-76}.
The first approach was extensively used
by Sawatzky and collaborators \cite{Veenendal93} . 
Here we use dipole transitions between single 
Slater determinants using non-orthogonal
orbitals, the orbitals of the final
state being relaxed in the presence
of the core hole. 

The possibility of describing the
electronic state of NiO by
a single Slater determinant was 
suggested in Refs ~\cite{Brandow-77,Brandow-92,Harrison-98}.
Moreover, relaxed Slater determinants can
sometimes describe a state much better than
the sum of a small number of unrelaxed Slater 
determinants~\cite{Brandow-77}.

A single Slater determinant is also the
non-interacting ground state of the Kohn-Sham
version of density funtional theory (DFT).
The corresponding Kohn-Sham orbitals are
usually considered to have no physical meaning.
This would be a problem for our approach
that calculates electric dipole transitions
between these orbitals. The success of DFT calculations
of XAS seems to indicate that Kohn-Sham orbitals
are physically meaningful and indeed
Gidopoulos~\cite{Gidopoulos-11} discovered that the non-interacting
Kohn-Sham ground state is the best approximation
of the true ground state in a subtle way.
To describe his finding, let
$h(\mathbf{r})=-\hbar^2\Delta/2m +v(\mathbf{r})$ be a one-body potential
and $H_v=\sum_i h(\mathbf{r}_i)$ be the corresponding
non-interacting many-body Hamiltonian. Denote by $|\Psi_v\rangle$
the (Slater determinant) ground state of $H_v$ and by $|\Psi\rangle$ the
ground state of the interacting Hamiltonian $H$.
By the Rayleigh-Ritz minimum principle we have
$\langle \Psi|H_v|\Psi\rangle - \langle \Psi_v|H_v|\Psi_v\rangle >0$.
Gidopoulos proved that the potential $v$ that minimizes
this difference is precisely the Kohn-Sham potential.
In that sense, the Kohn-Sham determinant and the
Kohn-Sham potential provide the best single-particle
description of the ground state of an interacting
system.

Therefore, it is relevant to describe
shake-up processes with non-orthogonal
Slater determinants. 
Other calculations were carried out
within this framework
by Tyson~\cite{TysonPhD},
who could calculate double-electron
excitations in XAS~\cite{Chaboy-94}
for LN$_{4,5}$-edges.
Similar calculations for x-ray photoemission
spectroscopy are more common~\cite{Takahashi-12}.

\subsection{Cross section}

The manybody X-ray photoemission cross section can be written as~\cite{almbladh}
\begin{eqnarray}
\sigma_{XPS}(\epsilon_k)&=&\frac{2\pi}{\hbar}\sum_f  |\langle k, \Psi_{f}
(N-1)|T|
\Phi_i(N)\rangle|^2 \nonumber \\ 
& & \delta(\epsilon_k-\hbar\omega- E_i(N)+E_f(N-1))
\label{eq:crossmbxps}
\end{eqnarray}
where $\epsilon_k$ is the photoelectron kinetic energy, $E_i(N)$ is the energy
of the $N$ electrons ground state $|\Phi_i(N)\rangle$, $E_f(N-1)$
and $|\Psi_{f}(N-1)\rangle$ characterize
the excited energy and state of the $N-1$ electron system with a
core hole
and $\hbar \omega$ is the energy of the
incident X-ray beam. The electric-dipole transition operator is
labeled T.
The transform $ I_{XPS}(t) $ of the XPS cross-section is defined as,
\begin{equation}
\sigma_{XPS}(\epsilon)=2 \Re \int_{0}^{+\infty} dt e^{i \epsilon_{+} t}
I_{XPS}(t)
\label{eq:Ixpst}
\end{equation} 
where $\epsilon_{+}=\epsilon+i \eta$ and Eq. \ref{eq:Ixpst} has to be understood
as the limit for $\eta\to 0^{+}$.

The manybody X-ray absorption cross section in the dipolar
approximation can be written as
\begin{eqnarray}
\sigma_{XAS}(\omega)&=&\frac{2\pi}{\hbar}\sum_f  |\langle \Psi_{f}(N)|W
|\Phi_i(N)\rangle|^2 \nonumber \\ 
& &\delta(E_f(N)-E_i(N)-\hbar \omega)
\label{eq:crossmbxas}
\end{eqnarray}
where now $\omega$ is the energy of the incident beam and
$W$ is proportional to the dipole transition operator $M$,
namely $W=\sqrt{2\pi\hbar^2\omega\alpha_0} M $. 

Similarly to what was done for the case of XPS and using 
a similar notation, we can define the 
the transform $ I_{XAS}(t) $ of the XAS cross-section as 
\begin{equation}
\sigma_{XAS}(t)=2 \Re \int_{0}^{+\infty} dt e^{i \omega_{+} t}
I_{XAS}(t)
\label{eq:Ixast}
\end{equation}  

Under the assumption that {\it both $\Phi_i$ and
  $\Psi_f^{XPS}$ are single determinant states}, 
Ohtaka and Tanake\cite{Ohtaka83,OhtakaRMP} demonstrated that
\begin{equation}
I_{XAS(t)}=I_{XPS}(t)  I_{0}(t)
\label{eq:OT}
\end{equation}
where both the terms $I_{XPS}(t)$ and $I_{0}(t)$ (see Eq. 4.43 in Ref. \cite{OhtakaRMP}) includes
manybody shake-up processes at all orders. Eq. \ref{eq:OT} holds for
a generic static core-hole potential. A similar relation was found for
the less-general case of a contact core-hole potential by Nozi\`eres and DeDominicis 
\cite{Nozieres69} using the linked cluster theorem.

If shake-up processes are neglected only in the $I_{0}(t)$ term then
$I_{0}(t)$ reduces to $I_{XAS}^{sp}(t)$, namely the transform of 
the single-particle XAS cross section calculated in the presence of a
static core-hole potential. Thus, it is possible to include to some
extent many body effects in the XAS cross section
$\sigma_{XAS}(\omega)$ 
by performing the convolution between the measured XPS cross section
$\sigma_{XPS}^{exp.}(\epsilon)$ that fully includes manybody shake-up
processes and the single-particle calculated XAS cross section
$\sigma_{XAS}^{sp}(\omega)$, namely
\begin{equation}
\sigma_{XAS}(\omega)=\int d\epsilon \, \sigma_{XPS}^{exp.}(\epsilon)
\sigma_{XAS}^{sp}(\omega-\epsilon)
\label{eq:convolution}
\end{equation}

In our work the single-particle cross section
$\sigma_{XAS}^{sp}(\omega) $
is calculated in the
framework of density functional theory with inclusion of a static 
core-hole and $ \sigma_{XPS}^{exp.}(\omega)$ is measured.

\subsection{Technical details}

The single particle XAS cross section $\sigma_{XAS}^{sp}(\omega) $ is
calculated in the framework of density functional theory using the implementation
of  Ref. \cite{Gougoussis09a} distributed with the Quantum-Espresso \cite{QE} distribution.
The technical details for the NiO calculation are the same as in 
ref. \cite{Gougoussis09b}. 
We used norm-conserving pseudopotentials with inclusion of semicore
states. The energy cutoffs used in the calculations were 140 Ryd and 160 Ryd
for CuO and CoO, respectively.
In the case of CoO, we neglect the tetragonal structural distortion
below the 290K N\'eel temperature and adopt magnetic and crystal
structures similar to those of NiO.
The electron-momentum grids for the Brillouin
zone integration and the choice of the supercell for the XAS
calculation are the same as for the NiO case in 
ref. \cite{Gougoussis09b}. 
The CuO XAS cross-section was calculated in the
supercell obtained by doubling the antiferromegnetic cell along the
shortest direction. 
The antiferromagnetic cell is obtained from the non-magnetic one by
defining as new lattice vectors ${\bf a}^{\prime}={\bf a}+{\bf c}$,
${\bf b}^{\prime}={\bf  b}$, and ${\bf c}^{\prime}={\bf a}-{\bf c}$
where ${\bf a}$,${\bf b}$, and ${\bf c}$ are the direct lattice vectors.
We then use a $3\times 3\times 3$ electron-momentum grid in the
supercell to obtain the self consistent charge density and a  
$3\times 3\times 3$ electron-momentum grid
in the supercell to calculate the XAS cross-section.

Finally we employ the DFT+U approximation in all case with  $U=7.75$ eV
and $U=11.1$ eV for CoO and CuO respectively. 
These values of the Hubbard repulsion are
calculated from first principles using the method of
ref. \cite{Cococcioni05}.

\section{Experiment}\label{sec:experiment}
The Ni-1s photoemission spectrum in NiO were measured at the HIKE
station of the KMC-1 beamline at BESSY
\cite{Gorgoi2009,Schaefers2007}. The spectra were recorded with a
SCIENTA R4000 photoelectron analyzer placed at 90$^\circ$ from the
x-ray beam. The incident x-ray beam (9 keV) was monochromatized by a pair of Si(422) crystals 
providing $\sim$500 meV energy bandwidth. 
To avoid charging effects, a 25 nm thick NiO thin film was grown on a
Ag substrate in the presence of oxygen, 
and capped by 3 nm of MgO. The growth of NiO was found fully epitaxial 
with the NiO(001) direction parallel to Ag(001) as confirmed by the
LEED patterns. 
The sample was positioned at a grazing angle of 89.99$^\circ$  from the incident x-rays in order to reduce the penetration depth of photons and enhance the photoelectron yield. 

The XAS spectra of NiO and CoO were borrowed from
Refs. \cite{Vedrinskii} and \cite{Modrow} respectively.

\section{Results}\label{sec:results}

\subsection{Nickel Oxide}

\begin{figure}[t]
\includegraphics[width=0.9\linewidth]{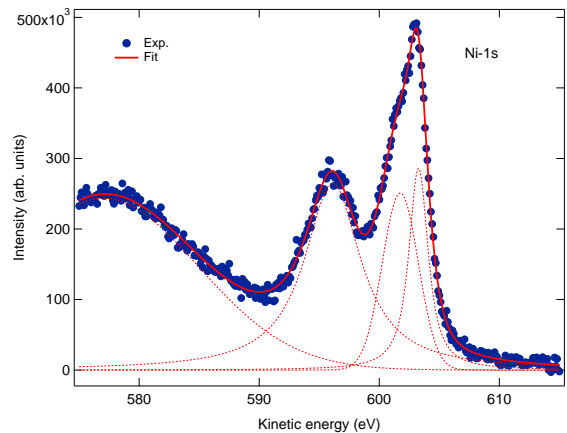}
\caption{figure 1: Experimental (circle) and fitted (lines) 1s photoemission spectra in NiO} 
\label{fig:NiOXPS}
\end{figure}

The measured 1s photoemission spectra of NiO are shown in
Fig. \ref{fig:NiOXPS}. The fit to the data is consistent with a three
peak structure in the 590-610 eV energy region. The results closely
resemble 2p$_{3/2}$ Ni photoemission in NiO
\cite{Taguchi,Parmigiani99} in this energy region.
In literature the attribution of the different features in 2p$_{3/2}$ Ni XPS is very
controversial and was subject to several reinterpretations. 
Van Veenendaal and Sawatzky \cite{Veenendal93} attributed the main feature at high energy
to a 2p$^5$3d$^9$\underline{L} state, where \underline{L} means a hole
in the ligand state. The satellite of the main peak (shoulder) was
attributed to non-local screening coming from the nearest neighbours
Ni atoms, while the lower energy satellite at $\approx 596 $ eV was
attributed to a 2p$^6$3d$^{10}$\underline{L}.
Recently this was reconsidered in ref. \cite{Taguchi} where the main 
feature was attributed to a 2p$^5$3d$^9$\underline{Z}, where 
\underline{Z} is a Zhang-Rice k-dispersing bound state\cite{Kunes}. The shoulder of the main
peak is attributed to a 2p$^5$3d$^9$\underline{L} state and the
lowest energy feature to a 2p$^5$3d$^8$ state.
Here we show that, regardless of their attribution, the features
measured in 1s Ni NiO XPS are present also in the dipolar Ni  K-edge
XAS spectrum
of NiO.
\begin{figure}[t]
\includegraphics[height=0.95\linewidth,angle=-90]{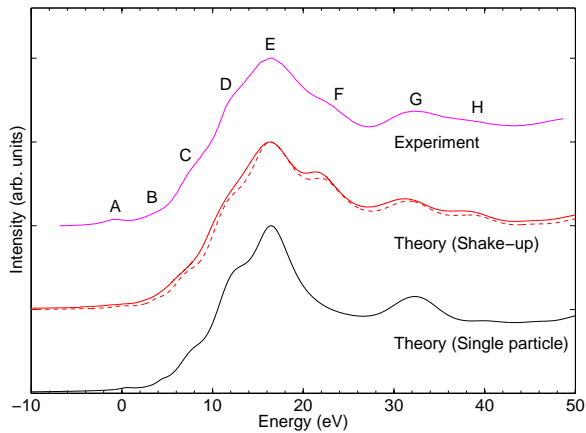}
\caption{Convolution of experimental XPS (this work) and single particle
XAS calculation in NiO at the Ni K-edge. 
Convolution with two or three component is shown with solid line and 
dashed line respectively (middle spectra). Experimental XAS data are from
Ref.\cite{Vedrinskii}.  }
\label{fig:NiOconv}
\end{figure}

In Fig. \ref{fig:NiOconv}  we show the measured and calculated XAS cross sections. The
single particle cross section is generally in good agreement with the 
dipolar part of the measured spectrum except for the two peaks indicated by
the letters F and H that are missing in the single particle calculation. 
In order to determine if the missing excitations are
manybody in nature, and eventually due to multi-determinant or shake-up processes, we then
proceed by using Eq. \ref{eq:convolution} and obtain new XAS spectra.
We first perform the convolution using the complete three-peaks structure
of the XPS spectra. We find that the convolution of the DFT calculated
XAS cross-section with the photoemission spectra greatly improves the
agreement with experiments. In particular the missing peaks are now
present in the spectrum and a better agreement occurs at all energy
scales. The F and H peaks are then replicas of the single particle
E and G peaks respectively and are manybody in nature.
We can further test to what extent this interpretation is robust by altering the XPS spectrum before convolution with the theoretical single-particle XAS calculation. We do this for NiO by artificially supressing the main peak of the XPS spectrum, leaving the shoulder and the satellite. We find that the resulting absorption spectrum (dashed line Fig xxx)is less in agreement with the experimental spectrum supporting our interpretation.

\subsection{Cobalt Oxide}

\begin{figure}[h]
\includegraphics[height=0.95\linewidth,angle=-90]{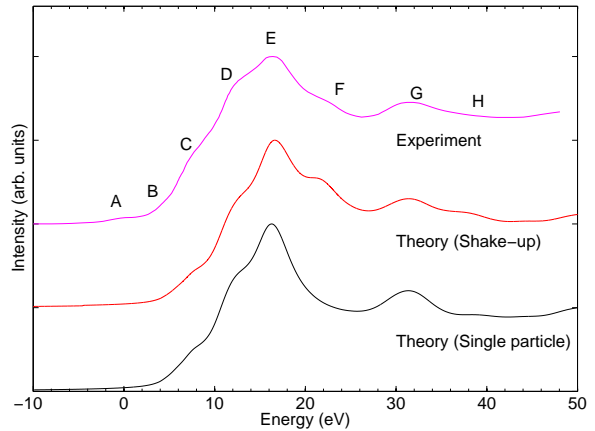}
\caption{Convolution of experimental XPS data of
  Refs. \cite{ShenCoO,Parmigiani99} 
and single particle XAS calculation in CoO at the Co K-edge. Experimental
XAS data are from Ref. \cite{Modrow}.}
\label{fig:CoOconv}
\end{figure}

Co 1s photoemission data of CoO are not available in literature.
However 2p$_{3/2}$and 3s Co XPS data \cite{ShenCoO,Parmigiani99}
are extremely similar and composed by two main peaks and a small
shoulder at low energy visible only in the 3s data. 
We then consider 3s photoemission data of Ref. \cite{Parmigiani99} 
and fit them with a 
two-peak structure and neglect the very small shoulder, 
invisible in 2p photoemission data. The results are shown in Fig. 
\ref{fig:CoOconv}.

The situation is very similar to NiO, namely the F peak is missing  
missing from the single-particle spectrum and the H peak is weak. Convolution with 
photoemission improves substantially the agreement although the main
edge peak is narrower than the experimental data.

\subsection{Copper Oxide}

The copper oxide CuO has a monoclinic crystal structure with symmetry
group C/2c .   The dipolar part in CuO was measured in ref. \cite{Bocharov01,BocharovTh}.
We follow the notation of refs. \cite{Bocharov01,BocharovTh} and label the configuration of the
crystal with respect to the incident beam by the three angles 
$(\theta, \phi, \psi)$. These angles correspond to the three rotation
angles of the goniometer. In particular, when the three angles are zero,
then the polarization is parallel to the $\theta$ axis of the goniometer 
and to the c-axis of the crystal. At zero angles, the plate holding the sample is orthogonal to
the incident beam and parallel to one of the plaquette chains in the
crystal (see Ref. \cite{BocharovTh} Fig. 10 for more details).

\begin{figure}[h]
\includegraphics[height=0.9\linewidth,angle=-90]{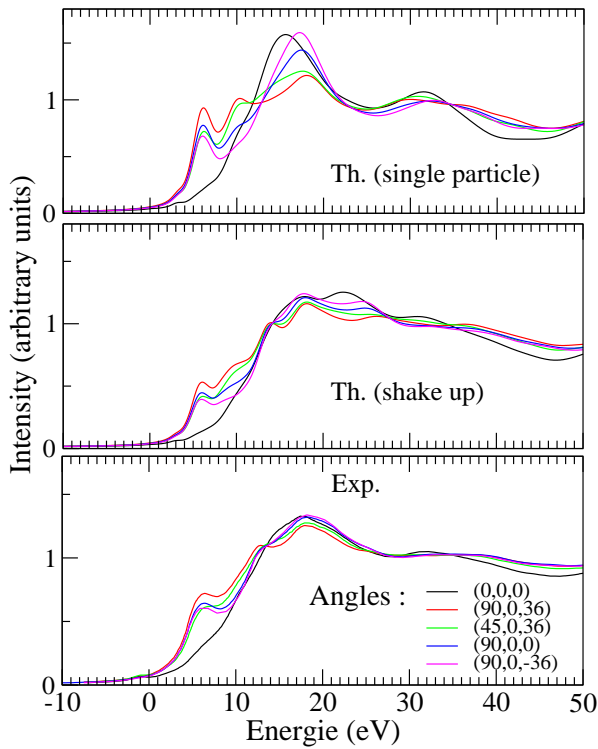}
\caption{Convolution between measured XPS from
  Ref. \cite{HiroshimaXPS} and
single particle XAS calculation at the Cu K-edge. Experimental XAS data are from
\cite{Bocharov01,BocharovTh}} 
\label{fig:cuo_conv}
\end{figure}

Given the low symmetry of the crystal, the polarization dependence of
CuO K-edge XAS spectra is very complicated, as can be seen in 
Fig. \ref{fig:cuo_conv}. The calculated single particle spectra are in
strong disagreement with experiments. Both the peak positions and the 
polarization dependence of the intensities disagree with the measured 
data. In order to see if  the disagreement is due to the lack of
manybody effects in the XAS cross-section, we consider the
convolution with Cu 1s photoemission spectra \cite{HiroshimaXPS}.
Cu 1s photoemission spectra of CuO are composed of two
peaks, usually attributed to 3d$^9$ and 3d$^{10}$\underline{L}.
Performing the convolution with the calculated single particle Cu K-edge XAS
leads to an impressive improvement. The convoluted spectrum is in much
better agreement with experimental data, demonstrating that the Cu
K-edge XAS spectrum in CuO is dominated by shake-up manybody
processes.

\section{Conclusions}

We have demonstrated that shake-up processes occur in  
dipolar K-edge XAS spectra of NiO, CoO, CuO. As these are prototype
correlated transition metal oxides, we expect these excitations
to be present in all XAS data of correlated materials. To be more
precise, whenever charge transfer satellites occur in XPS core-hole spectra, 
then shake-up satellites must also occur in the corresponding X-ray absorption
edge, at all energy scales.

We have also shown that a practical way to include these effects in first principle
calculations is to perform the convolution with the XPS spectrum at
the same edge, as suggested by Eq. \ref{eq:convolution}. Despite the
fact that Eq. \ref{eq:convolution} was obtained many years ago 
\cite{Nozieres69, Ohtaka83, OhtakaRMP}, we are currently unaware of 
other works trying to explicitly apply this equation to K-edge XAS by
using state of the art calculations. Our work that full includes
core-hole attraction and Hubbard U at the DFT+U level 
demonstrates that this approach is feasible and allows, for the 
first time, the attribution of all dipolar peaks in NiO and CoO
dipolar K-edge XAS spectra.

In Eq. \ref{eq:convolution} we neglected the additional many-body
terms \cite{Ohtaka83,OhtakaRMP} that are present in $I_0(t)$. These terms seem to be neglible
in NiO and CoO, but could explain the remaining discrepancy
between theory and experiment in CuO. Further work is required to 
calculate these many-body correction terms.

\section{Acknowledgement}

M. C. acknowledges fruitful discussion with F. Mauri. Calculations
were carried out at the IDRIS supercomputing center (proposal number:
091202).

\end{document}